\documentclass[12pt,a4paper]{article}
\usepackage{amsmath}
\usepackage{mitpress}

\newdimen\dummy
\dummy=\oddsidemargin
\usepackage{times}

\newlength{\wider}
\begin{document}


\noindent
{\LARGE \textbf{On the spectral distribution of photons
between 
\\[0.75ex]
planar interfaces}}
\\[2ex]
{\large\textbf{V. E. Mkrtchian}$^{1}$ and \textbf{C. Henkel}$^{2}$}
\\[1ex]
\normalsize
$^{1}$\textit{Institute for Physical Research, Armenian Academy of Sciences,
Ashtarak 0203, Republic of Armenia} 
\\[0.5ex] 
$^{2}$\textit{Institute of Physics and Astronomy, University of
Potsdam, Karl-Liebknecht-Str. 24/25, 14476 Potsdam, Germany} 


\begin{abstract}
\noindent Using a phenomenological approach to field quantization, an
expression for the Keldysh function of photons between two planar interfaces
(Casimir geometry) is found for any stationary quantum state of the two
bodies. The case of one interface sliding against the other is considered in
detail. \newline
\end{abstract}

\section{Introduction}

The simplest geometry in the electrodynamics of nanosystems is presumably
the Casimir system: two half-spaces of different materials separated by a
vacuum gap with two parallel planar interfaces. The gap has a width in the
nano-range and provides the stage for propagating and evanescent waves
emitted by the materials, thus shaping a rich plethora of Casimir physics:
Casimir pressure, quantum friction, near-field radiative heat transfer etc.
All of these phenomena share a purely electrodynamic origin, and the basic
theoretical quantity are the different components of Maxwell's stress
tensor. This tensor mirrors the quantum states of the electromagnetic field
in the gap, and the latter is determined by macroscopic properties of the
boundaries like their temperature distribution, current density and the
relative motion of the interfaces.

In the case of stationary macroscopic conditions, a general formula for the
distribution function of photons (Keldysh function) in the gap was recently
found in Ref.\cite{AdP}, based on tools from nonequilibrium quantum field
theory. This function provides a rather general description of the system
and contains quantities (called fluctuation sources) which are determined by
the above-mentioned macroscopic conditions of the system boundaries. In Ref.%
\cite{AdP}, the fluctuation sources were expressed by Keldysh functions
evaluated on the interfaces which are quite difficult to handle. In this
paper we find an alternative form for the sources that links them to the
number of excitations of the compound system electromagnetic field+bulk
medium, using the phenomenological theory of quantization of the
electromagnetic field in a dissipative medium~\cite{Jena}. We also give
explicit expressions for the Keldysh functions of a single interface system
for further applications.

To avoid overloading of the paper with formulae, our notation and
definitions follow those of Ref.\cite{AdP}.

\section{Preliminaries}

We consider two half-spaces of different media with parallel and homogeneous
boundaries (what we know about boundaries are theirs reflection
coefficients) located at $z = \nu a/2$ ($\nu = \pm$). The boundaries are in
stationary conditions: their temperatures are constant in time, and their
relative motion is uniform and parallel to each other. We can then assume
that the electromagnetic field (EMF) in the cavity [$-a/2\leq z\leq a/2$] is
stationary in time and statistically homogenous in the $xy$-plane. As a
consequence, all relevant fields and correlation functions can be expanded
in Fourier integrals with respect to frequency $\omega $ and wave vectors $%
\mathbf{q}=(q_{x},q_{y})$ along the interfaces. We use in the following the
shorthand $\Omega =(\Omega, \mathbf{q}).$ In the Dzyaloshinskii gauge where $%
\phi =0$, the transversality condition for the electric field can be used to
eliminate the normal components of the vector potential in favor of the
tangential ones (see \cite{AdP} for details) which are conveniently
expressed in the basis of s- and p-polarized fields of the lower interface
(index $\lambda =$ s, p).

The EMF in the gap is not a closed system because of dissipativity of the
enclosing boundaries, and the straightforward procedure of field
quantization in vacuum cannot be applied. So, for a quantum description of
the EMF in the gap, we have to resort to the theories of quantization in
dissipative media. There are many such theories~\cite%
{Jena,Agarwal75a,Rytov3,Barnett97}, but we shall take the simplest version
called phenomenological quantization~\cite{Jena} which recently found its
justification in the frame of the canonical quantization rules of quantum
field theory in the dissipative medium~\cite{Philbin}.

We begin with the splitting of the operator of vector potential $\mathbf{%
\hat{A}}$ into positive and negative parts in Fourier space%
\begin{eqnarray}
\mathbf{\hat{A}}(\mathbf{r},t) &=&\int \!\frac{d^{2}{q}}{(2\pi )^{2}}%
\int\limits_{0}^{+\infty }\!\frac{d\omega }{2\pi }\{\mathbf{\hat{A}}%
^{(+)}(\Omega ,z)e^{i(\mathbf{qr}-\omega t)}+\mathbf{\hat{A}}^{(-)}(\Omega
,z)e^{-i(\mathbf{qr}-\omega t)}\}\;;  \notag \\
&&\mathbf{\hat{A}}^{(-)}=[\mathbf{\hat{A}}^{(+)}]^{\dagger }  \label{eq:(1)}
\end{eqnarray}%
and then we postulate for the positive-frequency parts of the vector
potential a representation in terms of current sources 
\begin{equation}
\hat{A}_{\lambda }^{(+)}(\Omega ,z)=\sum_{\nu \lambda _{1}}D_{\lambda
\lambda _{1}}^{R}(\Omega ;z,\nu )\hat{I}_{\lambda _{1}}^{(+)}(\Omega ,\nu )
\label{eq:(2)}
\end{equation}%
at any point $z$ of the gap. 
The surface currents $\mathbf{\hat{I}}^{(+)}(\Omega ,\nu )$ here `live' on
the two interfaces $\nu =\pm $.

In Eq.(\ref{eq:(2)}), $\hat{D}^{R}$ is the retarded Green function (RGF) of
the EMF defined as ($\hbar=1$)%
\begin{equation}
D_{\lambda \lambda ^{\prime }}^{R}(\mathbf{r,}t;\mathbf{r}^{\prime
},t^{\prime })=-i\theta (t-t^{\prime })\left\langle [\hat{A}_{\lambda }(%
\mathbf{r,}t),\hat{A}_{\lambda ^{\prime }}(\mathbf{r}^{\prime }\mathbf{,}%
t^{\prime })]\right\rangle .  \label{eq:(3)}
\end{equation}%
The latter is a solution of an inhomogeneous wave equation with boundary
conditions. 
The intensity of the radiation is distributed according to the Keldysh Green
function (KGF) which is a quantum expectation value of a symmetrized field
correlation 
\begin{equation}
D_{\lambda \lambda ^{\prime }}^{K}(\mathbf{r,}t;\mathbf{r}^{\prime
},t^{\prime })=-i\left\langle \{\hat{A}_{\lambda }(\mathbf{r,}t),\hat{A}%
_{\lambda ^{\prime }}(\mathbf{r}^{\prime }\mathbf{,}t^{\prime
})\}\right\rangle .  \label{eq:(4)}
\end{equation}

To find RGF and KGF in Eqs.(\ref{eq:(3)}, \ref{eq:(4)}), we need averages of
the commutator and anticommutator of $\hat{A}_{\lambda }^{(+)}$ [Eq.(\ref%
{eq:(2)})] with its Hermitian conjugate $\hat{A}_{\lambda }^{(-)}$, and then
we come to compute averages of the commutator and anticommutator of the
surface currents $\hat{I}_{\lambda }^{(\pm )}$ introduced in Eq.(\ref{eq:(2)}%
). Our key assumption at this point is that we claim locality of the latter,
writing%
\begin{equation}
\big\langle\lbrack \hat{I}_{\lambda }^{(+)}(\Omega ,\nu ),\hat{I}_{\lambda
^{\prime }}^{(-)}(\Omega ^{\prime },\nu ^{\prime })]\big\rangle=\delta
(\Omega -\Omega ^{\prime })\delta _{\nu \nu ^{\prime }}c_{\lambda \lambda
^{\prime }}(\Omega ,\nu ),  \tag{5.a}
\end{equation}%
\begin{equation}
\big\langle\{\hat{I}_{\lambda }^{(+)}(\Omega ,\nu ),\hat{I}_{\lambda
^{\prime }}^{(-)}(\Omega ^{\prime },\nu ^{\prime })\}\big\rangle=\delta
(\Omega -\Omega ^{\prime })\delta _{\nu \nu ^{\prime }}a_{\lambda \lambda
^{\prime }}(\Omega ,\nu ).\stepcounter{equation}  \tag{5.b}
\end{equation}%
where we use the short-hand notation $\delta (\Omega )\equiv (2\pi
)^{3}\delta (\mathbf{q})\delta (\omega )$. Using Eqs.(5.a, \ref{eq:(2)}) and
the definition~(\ref{eq:(3)}) of RGF, we come an expression for the RGF that
has the same structure as Eq.(D9) of Ref.\cite{AdP}. Comparison of these two
expressions gives the relation%
\begin{equation}
\theta (\omega )\hat{c}(\Omega ,\nu )-\theta (-\omega )\hat{c}^{T}(-\Omega
,\nu )=i\hat{\Gamma}^{\nu }(\Omega )  \label{eq:(6)}
\end{equation}%
between the unknown matrix $\hat{c}$ in the r.h.s. of Eq.(5.a) and the
matrix $\hat{\Gamma}^{\nu }$ of Eq.(3.26) in Ref.\cite{AdP}. Taking into
account the properties of $\hat{\Gamma}^{\nu }(\Omega )$ (a symmetric matrix
and an odd function of $\Omega $), we find for $\hat{c}$ in Eq.(\ref{eq:(6)}%
) 
\begin{equation}
\hat{c}(\Omega ,\nu )=i\hat{\Gamma}^{\nu }(\Omega ).  \label{eq:(7)}
\end{equation}

With respect to the KGF~(\ref{eq:(4)}), using Eqs.(\ref{eq:(1)}, \ref{eq:(2)}%
), we can express it in the form of Eq.(4.21) of Ref.\cite{AdP}, where the
photon sources $\hat{P}(\Omega, \nu )$ are expressed by surface current
anticommutators (5.b) according to 
\begin{equation}
i\hat{P}(\Omega, \nu )=\theta (\omega )\hat{a}(\Omega, \nu )+\theta (-\omega
)\hat{a}^{T}(-\Omega, \nu ).  \label{eq:(8)}
\end{equation}
where $-\Omega = (-\omega, \mathbf{q})$. Using a quantization procedure, we
will show in the coming two sections that the photon sources $\hat{P}(
\Omega, \nu )$ \ref{eq:(8)} are proportional to $\hat{\Gamma}^{\nu }$
matrices as well: 
\begin{equation}
\hat{P}(\Omega, \nu ) = \mathcal{\hat{N}}_{\nu }^{f}(\Omega ) \hat{\Gamma}%
^{\nu}(\Omega ) \,,  \label{eq:(9)}
\end{equation}%
and the coefficients $\mathcal{\hat{N}}_{\nu }^{f}$ ($\nu =\pm $) of
proportionality are defined by excitation (occupation) numbers of the
compound system electromagnetic field+body below the interface at $z = \nu
a/2$.

\section{Quantization on the interface in rest}

There are two interfaces in the problem: the lower one is in at rest ($\nu
=- $) and the other one ($\nu =+$) is in a state of parallel uniform motion
in the $x$-direction, say. We begin our consideration with the lower
interface which is taken as a reference frame of the problem.

We introduce the annihilation operator $\hat{f}_{-,\lambda }$ of a Bose
field which is associated with the elementary excitations of the composed
system EMF+lower medium in the following way%
\begin{equation}
\hat{I}_{\lambda }^{(+)}(\Omega ,-)=\alpha _{\lambda \lambda _{1}}\left(
\Omega ,-\right) \hat{f}_{-,\lambda _{1}}(\Omega ).  \label{eq:(10)}
\end{equation}%
In Eq.(\ref{eq:(10)}), $\hat{\alpha}$ is an as yet unknown matrix, and we
suppose the canonical commutation rules for $\hat{f}_{-,\lambda }$%
\begin{equation}
\lbrack \hat{f}_{-,\lambda }(\Omega ),\hat{f}_{-,\lambda ^{\prime
}}^{\dagger }(\Omega ^{\prime })]=\delta _{\lambda \lambda ^{\prime }}\delta
(\Omega -\Omega ^{\prime }),  \tag{11.a}
\end{equation}%
\begin{equation}
\lbrack \hat{f}_{-,\lambda }(\Omega ),\hat{f}_{-,\lambda ^{\prime }}(\Omega
^{\prime })]=[\hat{f}_{-,\lambda }^{\dagger }(\Omega ),\hat{f}_{-,\lambda
^{\prime }}^{\dagger }(\Omega ^{\prime })]=0.\stepcounter{equation} 
\tag{11.b}
\end{equation}%
%
%
%
We restrict our consideration considering a quantum state for the lower
surface which is diagonal in the chosen basis and characterized by the
excitation number $N_{-,\lambda }^{f}(\Omega )$ 
\begin{equation}
\big\langle\hat{f}_{-,\lambda }^{\dagger }(\Omega )\hat{f}_{-,\lambda
^{\prime }}(\Omega ^{\prime })\big\rangle=\delta _{\lambda \lambda ^{\prime
}}\delta (\Omega -\Omega ^{\prime })N_{-,\lambda }^{f}(\Omega ).
\label{eq:(12)}
\end{equation}

Then, for the nonzero averages of commutator and anticommutator~(5) of the
surface currents on the lower interface, we have correspondingly%
\begin{equation}
\hat{c}(\Omega ,-)=\hat{\alpha}\left( \Omega ,-\right) \hat{\alpha}^{\dagger
}\left( \Omega ,-\right)  \tag{13.a}
\end{equation}%
\begin{equation}
\hat{a}(\Omega ,-)=\hat{\alpha}\left( \Omega ,-\right) (\hat{I}+2\hat{N}%
_{-}^{f})\hat{\alpha}^{\dagger }\left( \Omega ,-\right) \stepcounter{equation%
}  \tag{13.b}
\end{equation}%
where $\hat{I}$ is a $2\times 2$ unit matrix and we introduced the diagonal
matrix 
\begin{equation}
\hat{N}_{-}^{f}\equiv \left( 
\begin{array}{cc}
N_{-,s}^{f} & 0 \\ 
0 & N_{-,p}^{f}%
\end{array}%
\right)  \label{eq:(14)}
\end{equation}%
Taking into account that $\hat{\Gamma}^{-}$ in Eq.(\ref{eq:(7)}) is
diagonal, we also suggest the diagonality of $\hat{\alpha}\left( \Omega
,-\right) $ in (13.a). This provides the anticommutator (13.b) and using Eq.(%
\ref{eq:(12)}), we find%
\begin{equation}
\hat{a}(\Omega ,-)=i[\hat{I}+2\hat{N}_{-}^{f}\left( \Omega \right) ]\hat{%
\Gamma}^{-}(\Omega )  \label{eq:(15)}
\end{equation}%
Insertion of Eqs.(\ref{eq:(15)}, 13.b) into Eq.(\ref{eq:(8)}) gives us
expression~(\ref{eq:(9)}) for $\nu =-$ where $\mathcal{\hat{N}}_{-}^{f}$ is
defined in the same way as the photon number in free space [see Eq.(C4) of
Ref.\cite{AdP}] 
\begin{equation}
\mathcal{\hat{N}}_{-}^{f}(\Omega )=\hat{I}\mathop{\rm sign}\omega +2[\theta
(\omega )\hat{N}_{-}^{f}(\Omega )-\theta (-\omega )\hat{N}_{-}^{f}(-\Omega )]
\label{eq:(16)}
\end{equation}

\section{Quantization on the sliding interface}

Using the transformation law for the surface currents and the properties of
Fourier transforms under the Lorentz transformation of the space-time
coordinates, we find the positive-frequency part of the surface current
operator in the laboratory frame $\hat{I}_{\lambda }^{(+)}(\Omega ,+)$. It
is expressed via corresponding operators in the reference frame of the
moving interface [see Eq.(D2) of Ref.\cite{AdP}] 
\begin{equation}
\hat{I}_{\lambda }^{(+)}(\Omega ,+)=O_{\lambda \lambda ^{\prime }}(\Omega
)[\theta (\omega ^{\prime })\hat{I}_{\lambda ^{\prime }}^{\prime (+)}(\Omega
^{\prime },+)+\theta (-\omega ^{\prime })\hat{I}_{\lambda ^{\prime
}}^{\prime (-)}(-\Omega ^{\prime },+)];\text{ }\omega \geq 0  \label{eq:(17)}
\end{equation}%
where $\Omega ^{\prime }=(\omega ^{\prime },q_{x}^{\prime },q_{y})$ is
related to the reference frame $K^{\prime }$ co-moving with the upper
interface; it is connected with $\Omega =(\Omega ,q_{x},q_{y})$ via a
Lorentz transformation. The transformation matrix $\hat{O}(\Omega )$ in Eq.(%
\ref{eq:(17)}) is given in Eq.(D4) of Ref.\cite{AdP}.

Introducing a Bose field (with annihilation operator $\hat{f}_{+,\lambda
^{\prime }}^{\prime }$) of excitations in the rest frame of the upper body
analogous to Eqs.(\ref{eq:(10)}, 11) 
\begin{equation}
\hat{I}_{\lambda ^{\prime }}^{\prime (+)}(\Omega ^{\prime },+)=\alpha
_{\lambda ^{\prime }\lambda _{1}^{\prime }}^{\prime }\left( \Omega ^{\prime
},+\right) \hat{f}_{+,\lambda _{1}^{\prime }}^{\prime }(\Omega ^{\prime }) 
\tag{18.a}
\end{equation}%
we find for the surface current anticommutator in $K^{\prime }$%
\begin{equation}
\big\langle\{\hat{I}_{\lambda _{1}^{\prime }}^{(+)}(\Omega _{1}^{\prime },+),%
\hat{I}_{\lambda _{2}^{\prime }}^{(-)}(\Omega _{2}^{\prime },+)\}\big\rangle%
=i\delta (\Omega _{1}^{\prime }-\Omega _{2}^{\prime })[I+2\hat{N}%
_{+}^{\prime f}(\Omega _{1}^{\prime })]\hat{\Gamma}^{+\prime }(\Omega
_{1}^{\prime })\stepcounter{equation}  \tag{18.b}
\end{equation}%
And finally, using expression (17) for the positive-frequency part of the
current operator in $K$, we find the anticommutator in the laboratory frame
for $\omega \geq 0$%
\begin{equation}
\hat{a}(\Omega ,+)=i\hat{O}\mathcal{\hat{N}}_{+}^{\prime f}(\Omega ^{\prime
})\hat{\Gamma}^{\prime +}(\Omega ^{\prime })\hat{O}^{T}  \tag{19.a}
\end{equation}%
where%
\begin{equation}
\mathcal{\hat{N}}_{+}^{\prime f}(\Omega ^{\prime })=\hat{I}\mathop{\rm sign}\omega
^{\prime }+2[\theta (\omega ^{\prime })\hat{N}_{+}^{\prime f}(\Omega
^{\prime })-\theta (-\omega ^{\prime })\hat{N}_{+}^{\prime f}(-\Omega
^{\prime })].\stepcounter{equation}  \tag{19.b}
\end{equation}

Using Eq.(\ref{eq:(8)}) for $\nu =+$ we find for the photon sources $\hat{P}%
(\Omega, +)$ on the moving interface the expression%
\begin{equation}
\hat{P}(\Omega, +)=\hat{O}\mathcal{\hat{N}}_{+}^{\prime f}(\Omega ^{\prime })%
\hat{\Gamma}^{\prime +}(\Omega ^{\prime })\hat{O}^{T}  \label{eq:(20)}
\end{equation}
Taking into account the transformation law of $\hat{\Gamma}^{+}$ [Eq.(D11)
of Ref.\cite{AdP}], we come to the expression~(\ref{eq:(9)}) for $\nu =+$
where the excitation number $\mathcal{\hat{N}}_{+}^{f}$ in the laboratory
frame $K$ is given by%
\begin{equation}
\mathcal{\hat{N}}_{+}^{f}(\Omega ) = \hat{O}\mathcal{\hat{N}}_{+}^{\prime
f}(\Omega ^{\prime })\hat{O}^{-1}.  \label{eq:(21)}
\end{equation}
Because of the invariance of the trace of the matrix under the similarity
transformation in Eq.(\ref{eq:(21)}), the number of excitations in $K$ and $%
K^{\prime }$ are the same, and the relative motion only changes their
polarization. 

Obviously Eq.(\ref{eq:(20)}) recovers the result~(7.3) of Ref.\cite{AdP} in
the case of thermal equilibrium in $K^{\prime }$.

\section{Examples of Keldysh Green functions}

In this paragraph we give expressions for the KGF of Casimir system and for
the single interface system for further applications.

The expression for KGF of the Casimir system with planar parallel boundaries
is already written in Ref.\cite{AdP}, Eqs.(6.7--6.11). This is cumbersome
enough not to repeat it here. We only give explicitly the matrices $\hat{%
\gamma}_{\nu }$ defined in Eq.(6.11) of Ref.\cite{AdP} which characterize
the fluctuation sources in KGF of the Casimir system.

Insertion of Eq.(\ref{eq:(9)}) into the definition of $\hat{\gamma}_{\nu}$,
we find after some algebra 
\begin{equation}
\hat{\gamma}_{-}=e^{-a\mathop{\rm Im}q_{z}}\mathcal{\hat{N}}_{-}^{f}\mathcal{\hat{R%
}}_{-}\hat{\Delta}_{0}  \tag{22.a}
\end{equation}%
for the sources in the lower interface and%
\begin{equation}
\hat{\gamma}_{+}=e^{-a\mathop{\rm Im}q_{z}}(\hat{I} + \hat{R}_{+})\hat{\Delta}_{0}%
\mathcal{\hat{N}}_{+}^{f}\hat{\Delta}_{0}^{-1}(\hat{I} + \hat{R}_{+})^{-1}%
\mathcal{\hat{R}}_{+}\hat{\Delta}_{0} \stepcounter{equation}  \tag{22.b}
\end{equation}%
in the upper, moving one. Here, we have put 
%
\begin{equation}
\mathcal{\hat{R}}_{\nu } = \left\{ 
\begin{array}{ll}
\hat{I} - \hat{R}_{\nu }\hat{R}_{\nu }^{\ast } & \quad\text{ for propagating
waves} \\ 
2i\mathop{\rm Im}\hat{R}_{\nu } & \quad\text{ for evanescent waves}%
\end{array}%
\right.  \label{eq:(23)}
\end{equation}
where propagating (evanescent) waves are defined by the observer-independent
inequality $q_z^2 = (\omega/c)^2 - {q}^2 > 0$ ($q_z^2 < 0$).

We get the KGF for a single, moving body (located in $z\geq 0$) by applying
a limiting procedure where the two points $z,z^{\prime }\leq 0$ remain
fixed, while the lower interface recedes to infinity (limit called (\textbf{C%
}) in Ref.\cite{AdP}). This yields 
\begin{align}
\hat{D}_{+}^{K}(\Omega ;z,z^{\prime })=& \{(\hat{I}+\hat{R}_{+})\hat{\Delta}%
_{0}\mathcal{\hat{N}}_{+}^{f}\hat{\Delta}_{0}^{-1}(\hat{I}+\hat{R}_{+})^{-1}%
\mathcal{\hat{R}}_{+}e^{-i(q_{z}z-q_{z}^{\ast }z^{\prime })}+  \tag{24.a} \\
& {}+\theta (q_{z}^{2})[\hat{I}\,e^{iq_{z}z}+\hat{R}_{+}e^{-iq_{z}z}]%
\mathcal{\hat{N}}_{-}^{f}[\hat{I}\,e^{-iq_{z}z^{\prime }}+\hat{R}_{+}^{\ast
}e^{iq_{z}z^{\prime }}]\}\hat{\Delta}_{0}  \notag
\end{align}%
%
%
where $\mathcal{\hat{N}}_{-}^{f}(\Omega )$ is now interpreted as the average
number of `up-propagating' photons that are incident on the moving body. 
If the upper interface is at rest, all matrices in (24.a) are diagonal, and
we arrive at 
\begin{align}
\hat{D}_{+}^{K}(\Omega ;z,z^{\prime })=& \{\mathcal{\hat{N}}_{+}^{f}\mathcal{%
\hat{R}}_{+}e^{-i(q_{z}z-q_{z}^{\ast }z^{\prime })}+  \tag{24.b} \\
& {}+\theta (q_{z}^{2})[\hat{I}\,e^{iq_{z}z}+\hat{R}_{+}e^{-iq_{z}z}]%
\mathcal{\hat{N}}_{-}^{f}[\hat{I}\,e^{-iq_{z}z^{\prime }}+\hat{R}_{+}^{\ast
}e^{iq_{z}z^{\prime }}]\}\hat{\Delta}_{0}  \notag
\end{align}%
Finally, letting both bodies recede to infinity, with $z$, $z^{\prime }$
kept finite [limit(\textbf{A}) of Ref.\cite{AdP}], we find the KGF $\hat{D}%
_{0}^{K}$ of the EMF in free space 
\begin{equation}
\hat{D}_{0}^{K}(\Omega ;z,z^{\prime })=\theta (q_{z}^{2})[\mathcal{\hat{N}}%
_{+}e^{-iq_{z}(z-z^{\prime })}+\mathcal{\hat{N}}_{-}e^{iq_{z}(z-z^{\prime
})}]\hat{\Delta}_{0}\stepcounter{equation}  \tag{24.c}
\end{equation}
In the expressions (24), $\mathcal{\hat{N}}_{\pm }$ are defining the number
of free photons moving in opposite directions, which can be found in free
space also by elementary plane wave quantization [Eq.(C4) of Ref.\cite{AdP}].

\section{Summary}

We complete our considerations~ \cite{AdP} of the photonic Keldysh function
in the Casimir geometry of two plates by deriving explicit expressions for
the fluctuation sources using phenomenological quantization of the
electromagnetic field in a dissipative medium. We find in particular simple
expressions for the Keldysh functions for a single interface system in an
arbitrary stationary non-equilibrium state.

\section{Acknowledgments}

We acknowledge financial support by the European Science Foundation (ESF)
within the activity \textquotedblleft New Trends and Applications of the
Casimir Effect\textquotedblright{} (exchange grant 2847), and by the
Deutsche Forschungsgemeinschaft (grant He 2849/4-1).


\end{document}